\documentstyle[prl,aps,preprint,amsfonts]{revtex}

\begin{document}
\title{AN INFRARED SINGULARITY IN THE DAMPING RATE\\
FOR LONGITUDINAL GLUONS IN HOT QCD}
\author{A. Abada and O. Azi}
\address{D\'epartement de Physique, \'Ecole Normale Sup\'erieure\\
BP 92 Vieux Kouba, 16050 Alger, Algeria\\
{\tt enskppps@ist.cerist.dz}}
\date{\today}
\maketitle  
\begin{abstract}
We calculate $\gamma _l(0)$, the damping rate for longitudinal on-shell
gluons with zero momentum in hot QCD using the hard-thermal-loop (htl)
scheme. We find it to be divergent in the infrared, which means that in this
scheme $\gamma _l(0)$ is different from $\gamma _t(0)$, the corresponding
damping rate for transverse gluons which is known to be finite. This result
suggests that the htl scheme is infrared sensitive and thus may need to be
improved upon in this sector. We discuss this issue after we present our
calculation.
\end{abstract}

\bigskip

{\small pacs: 11.10.Wx 12.38.-t 12.38.Bx 12.38.Mh}

{\small keywords: hard thermal loops. soft gluon damping.}

ENSK--TP--10

\vspace{0.3in}

Besides their importance regarding the stability of the quark-gluon plasma,
gluon damping rates have been crucial in better understanding QCD at high
temperature $T$. In early works, it has been noticed that in this regime,
the determination of the dispersion laws for quarks and gluons beyond lowest
order using standard perturbation theory is gauge-dependent \cite{early}.
This problem has been emphasized in further works in which the gluon damping
rates have been calculated to one-loop order in various gauges and schemes
and different results have been obtained\cite{later}. It was then realized
that the problem was related to the way the expansion in powers of $g$, the
perturbative QCD\ coupling constant, was performed: at high $T$ (the hard
scale), when the external momenta are soft, i.e., of magnitude $gT,$ the
standard loop expansion is not anymore an expansion in powers of $g$ \cite
{Pisarski1}. It was subsequently developed an effective perturbative
expansion in the framework of a resummation scheme of the so-called hard
thermal loops (htl) \cite{BPhtl}. Using this scheme, the transverse-gluon
damping rate $\gamma _t(0)$ with zero momentum was shown to be
Coulomb-and-covariant-gauge-invariant and determined in the strict Coulomb
gauge to be finite and positive \cite{BPgamt}. Later, a
generating-functional formalism in the htl approximation was developed \cite
{action} and a relation to the eikonal of a Chern-Simons gauge theory was
made \cite{Chern-Simons}. From there a hydrodynamic approach showed that the
htl approximation was essentially `classical' \cite{hydro}.

However, the htl resummation scheme may be not `complete' yet in describing
the whole picture of QCD at high temperature. Indeed, it discusses only the
two scales $T$ (hard)\ and $gT$ (soft) whereas with $g$ and $T,$ one has (at
least)\ a hierarchy of scales $g^nT$, $n$ an integer, positive if we believe
that $T$ is the highest scale in the problem. It may be unnecessary to
reorganize perturbation theory taking into account this general structure of
scales, but there are indications that at least\ the next smaller one $g^2T$
may be important. Indeed, it is true that in the htl scheme, both on-shell
longitudinal and transverse gluons acquire a thermal mass $m_g$ of order $gT$%
, but static magnetic fields are not screened yet at this scale \cite{magne}%
; they are expected to get so at the hence-called `magnetic scale' $g^2T$.
Furthermore, the gluon damping rates $\gamma (p)$ are to lowest order of
magnitude $g^2T$ and it has been noticed that these damping rates, when
calculated in the htl-resummation scheme at soft but nonzero momenta $p$,
exhibit a $\ln g$ behavior accompanied with coefficients \cite{logdiv}. This
logarithmic behavior is not present in the expression of $\gamma _t(0)$, the
limit $p\rightarrow 0$ of $\gamma _t(p),$ and no explanation has been
provided yet for this discrepancy. These remarks tend to indicate therefore
that the htl scheme may be sensitive in the `infrared'.

We think a more direct indication of the infrared sensitivity of the
htl-resummation scheme is the observation we made in \cite{AAB} that, when
calculated solely in this scheme, the longitudinal-gluon damping rate at
zero momentum $\gamma _l(0)$ may be potentially infrared-divergent. Indeed,
we have determined in that work the analytic expression to leading order of $%
\gamma _l(0)$. We have obtained for it the following form: 
\begin{equation}
\gamma _l(0)=\frac{g^2N_cT}{24\pi }\left[ a_{l0}^{(1)}+a_{l0}^{(2)}\right]
\,;  \label{gamml}
\end{equation}
$N_c$ is the number of colors. $a_{l0}^{(1)}$ is given in eq (\ref{al01})
below; it is a finite number found in \cite{BPgamt} to be equal to 6.63538\
and is such that the damping rate $\gamma _t(0)$ for transverse gluons with
zero momentum is equal to $\frac{g^2N_cT}{24\pi }\,a_{l0}^{(1)}$, see \cite
{BPgamt}. The other contribution $a_{l0}^{(2)}$ is given in eq (\ref{al02})
and it is this part of $\gamma _l(0)$ we remarked it contains terms which
are divergent in the infrared \cite{AAB}.

From a physical standpoint and independently of any calculation scheme,
there is no difference at zero momentum between longitudinal and transverse
gluons, and hence one expects any consistent scheme to yield the same result
for both damping rates \cite{BPgamt}. In virtue of this physical argument,
once $\gamma _t(0)$ was found in the htl scheme to be finite and positive, $%
\gamma _l(0)$ was thought to be so and to the best of our knowledge, no
explicit calculation of the latter in the htl scheme has been reported prior
to the one in \cite{AAB}. With this in mind and expecting the
htl-resummation scheme to be `complete', we argued in \cite{AAB} that the
terms contributing to $a_{l0}^{(2)}$, in particular the infrared-divergent
ones, when put together, may cancel one another. We show in this letter that
this is not the case: when all the expressions are treated with care, not
only we find the part $a_{l0}^{(2)}$ nonvanishing but the infrared-divergent
piece survives too, see eq (\ref{final}). In the sequel, we present first
our calculation and we discuss the result afterwards.

The longitudinal gluon damping rate in the strict Coulomb gauge is obtained
to lowest order by the relation: 
\begin{equation}
\gamma _l(p)=\frac{{\rm Im}\,^{*}\Pi _l(-i\omega ,p)}{\frac \partial {%
\partial \omega }\delta \Pi _l(-i\omega ,p)}\Big | _{\omega =\omega
_l(p)+i0^{+}}\,,  \label{def-gaml}
\end{equation}
where $\delta \Pi _l$ ($^{*}\Pi _l$) is the longitudinal-gluon htl
(effective) self-energy. $\omega _l(p)$ is its on-shell lowest-order energy
given for soft momenta in units of $m_g$ by eq (\ref{omegatl}) below. We
follow closely the notation of \cite{AAB} to which this work is a follow-up.
Since the denominator in (\ref{def-gaml}) is given in units of $m_g$ by $%
\partial /\partial \omega \,\delta \Pi (-i\omega ,p)|_{\omega =\omega
_l(p)}=-2p^2+...\,$, to get $\gamma _l(p=0)$, we have to expand the
imaginary part of the effective self-energy to order $p^2$. This calculation
is reported in \cite{AAB} and we obtain the result (\ref{gamml}) where the
expressions of $a_{l0}^{(1)}$ and $a_{l0}^{(2)}$ are the following: 
\begin{eqnarray}
a_{l0}^{(1)} &=&9\int_0^{+\infty }dk\int_{-\infty }^{+\infty }{\frac{d\omega
_1}{\omega _1}}\int_{-\infty }^{+\infty }{\frac{d\omega _2}{\omega _2}}%
\delta (1-\omega _1-\omega _2)\big[\,k^4\,\rho _{l1}\,\rho
_{l2}-k^2(k^2-\omega _1^2)^2  \nonumber \\
&&\ \ \ \ \ \times \,\rho _{t1}\rho _{l2}+2(k^2+\omega _1\omega _2)^2\,\rho
_{t1}\,\rho _{t2}+{\frac{\omega _1}{6k^3}(}k^2-\omega _1^2)^2\Theta _1\,\rho
_{t2}\big]\,,  \label{al01}
\end{eqnarray}
and: 
\begin{eqnarray}
a_{l0}^{(2)} &=&9\int_0^{+\infty }dk\int_{-\infty }^{+\infty }{\frac{d\omega
_1}{\omega _1}}\int_{-\infty }^{+\infty }{\frac{d\omega _2}{\omega _2}}%
\,\delta (1-\omega _1-\omega _2)\Big[{\frac{\omega _1}3}(1-2\omega _1)\Theta
_1\,\partial _k\rho _{l2}+{\frac{\omega _1k}6}\Theta _1\,\partial _k^2\rho
_{l2}  \nonumber \\
&&\ \ +{\frac{\omega _1}3}\Big( 1-{\frac{\omega _1^2}{k^2}}\Big) (1-2\omega
_1)\partial _k\Theta _1\,\rho _{t2}+{\frac{\omega _1k}3}\Big( 1-{\frac{%
\omega _1^2}{k^2}}\Big) \,\partial _k\Theta _1\,\partial _k\rho _{t2}+{\frac{%
\omega _1}{6k}}\big( k^2-\omega _1^2\big) \partial _k^2\Theta _1\,\rho _{t2}
\nonumber \\
&&\ \ +\partial _k\Big[{k}^3\Big( 1+{\frac{k^2}3}\Big) \,\rho _{l1}\rho
_{l2}+{\frac{2k}9}\big(k^2(1-3k^2)+(1-4k^2)\omega _1\omega _2-\omega
_1^2\omega _2^2\big)\rho _{t1}\rho _{t2}  \nonumber \\
&&\ \ +{\frac{2\omega _1^2}3}\Big( 1+{\frac{\omega _1}{k^2}}-{\frac{\omega
_1^2}{k^2}}\Big) \Theta _1\,\rho _{t2}+{\frac{k^2}9}\big( %
1-2k^2-4(1-k^2)\omega _1+6\omega _1^2-4\omega _1^3\big) \rho _{t1}\,\partial
_k\rho _{t2}  \nonumber \\
&&\ \ -{\frac{\omega _1}6}\partial _k\Big( \Big( k-{\frac{\omega _1^2}k}%
\Big) \Theta _1\,\rho _{t2}\Big) \Big] +\frac 23|\omega _1|({2}k^2\delta
_1\rho _{l2}-k^2\omega _1\partial _{\omega _1^2}\delta _1\,\rho _{l2}+\omega
_1\delta _1\rho _{t2})\Big].  \label{al02}
\end{eqnarray}
The notation is as follows: $\Theta _1\equiv \Theta (k^2-\omega _1^2)$ where 
$\Theta $ is the step function and $\delta _1\equiv \delta (k^2-\omega _1^2)$%
. $\partial _k\equiv \partial /\partial k$ etc and $\rho _{li}$ is a short
notation for $\rho _l(\omega _i,k),\,i=1,2$ and the same for $\rho _{ti}$.
The spectral densities $\rho _{t,l}$ are given by \cite{BPgamt,rho}: 
\begin{equation}
\rho _{t,l}(\omega ,k)=\,{\frak z}_{t,l}(k)\left[ \delta \left( \omega
-\omega _{t,l}(k)\right) -\delta \left( \omega +\omega _{t,l}(k)\right)
\right] +\beta _{t,l}(\omega ,k)\,\Theta (k^2-\omega ^2)\,,  \label{rhos}
\end{equation}
an expression in which the residues $\,{\frak z}_{t,l}(k)$ are given by: 
\begin{equation}
\,{\frak z}_t(k)=\,\frac{(\omega ^2-k^2)}{2(3\omega ^2-(\omega ^2-k^2)^2)}%
\Big | _{\omega =\omega _t(k)}\,;\quad \,{\frak z}_l(k)=\frac{-(\omega
^2-k^2)}{2k^2(3-\omega ^2+k^2)}\Big | _{\omega =\omega _l(k)}\,,  \label{ztl}
\end{equation}
and the cut functions $\beta _{t,l}(\omega ,k)$ by: 
\begin{eqnarray}
\beta _t(\omega ,k) &=&\frac{3\omega (k^2-\omega ^2)}{4k^3[(k^2-\omega ^2+%
\frac{3\omega ^2}{2k^2}(1+\frac{k^2-\omega ^2}{2\omega k}\ln \frac{k+\omega 
}{k-\omega }))^2+(\frac{3\pi \omega }{4k^3}(k^2-\omega ^2))^2]}\,;  \nonumber
\\
\beta _l(\omega ,k) &=&\frac{-3\omega }{2k[(3+k^2-\frac{3\omega }{2k}\ln 
\frac{k+\omega }{k-\omega })^2+(\frac{3\pi \omega }{2k})^2]}\,.
\label{betatl}
\end{eqnarray}
$\omega _{t,l}(p)$ are the on-shell energies of the gluon to lowest order.
For soft gluons and in units of $m_g$, we have: 
\begin{eqnarray}
\omega _t(p) &=&\Big[1+{\frac 35}\,p^2-{\frac 9{35}}\,p^4+{\frac{704}{3000}}%
\,p^6-{\frac{91617}{336875}}\,p^8+\dots \Big]\ ;  \nonumber \\
\omega _l(p) &=&\Big[1+{\frac 3{10}}\,p^2-{\frac 3{280}}\,p^4+{\frac 1{6000}}%
\,p^6+{\frac{489}{43120000}}\,p^8+\dots \Big]\,.  \label{omegatl}
\end{eqnarray}

As we said, $a_{l0}^{(1)}$ is a finite number found in \cite{BPgamt} to be
equal to 6.63538. We therefore have to calculate $a_{l0}^{(2)}$. The sort of
difficulties one encounters when dealing with the expressions involved in eq
(\ref{al02}) can partly be displayed in the following simple example.
Consider the integral:

\begin{eqnarray}
I &=&12\int_0^{+\infty }dk\int_{-\infty }^{+\infty }{\frac{d\omega _1}{%
\omega _1}}\int_{-\infty }^{+\infty }{\frac{d\omega _2}{\omega _2}}\,\delta
(1-\omega _1-\omega _2)\,k^2|\omega _1|\,{\frak z}_l(k)\,\delta (k^2-\omega
_1^2)\,\delta (\omega _2-\omega _l(k))  \nonumber \\
\ &=&12\int_0^{+\infty }dk\int_{-\infty }^{+\infty }{d\omega \,}\frac{\,%
{\frak z}_l(k)\,k^2\epsilon (\omega )}{1-\omega }\,\delta (k^2-\omega
^2)\,\delta (1-\omega -\omega _l(k))\,,  \label{example1}
\end{eqnarray}
the residue piece of a term that intervenes in the expression of $%
a_{l0}^{(2)}$. Here $\epsilon (\omega )$ is the sign function. We may choose
to write $\delta (k^2-\omega ^2)=\frac 1{2k}\delta (k-|\omega |)\,,$ in
which case we obtain: 
\begin{eqnarray}
I &=&-6\int_0^{+\infty }dk{\,}\frac{k\,\,{\frak z}_l(k)}{\omega _l(k)}%
\,\delta (k+1-\omega _l(k))=-6\,\frac{k\,\,{\frak z}_l(k)}{\omega _l(k)}%
\,\left| \frac{d(1+k-\omega _l(k))}{dk}\right| ^{-1}\Big|_{k\rightarrow 0} 
\nonumber \\
\ &=&\frac 3{2\eta }+\frac 9{10}+O(\eta )\,.  \label{example2}
\end{eqnarray}
We have used the fact that $\omega _l(k)\geq 1$ for all $k$ and an expansion
of ${\frak z}_l(k)$ and $\omega _l(k)$ for small $k$. In the above equation, 
$\eta $ is an infrared cutoff; $\eta \ll 1.$ Thus $I$ is divergent in the
infrared. But we may also choose to write $\delta (k^2-\omega ^2)=\frac 1{%
2|\omega |}\delta (k-|\omega |)\,,$ in which case we have: 
\begin{eqnarray}
I &=&6\int_0^{+\infty }dk{\,}\frac{k^2\,\,{\frak z}_l(k)}{\omega
_l(k)(1-\omega _l(k))}\,\delta (k+1-\omega _l(k))  \nonumber \\
\ &=&6\,\frac{k^2\,\,{\frak z}_l(k)}{\omega _l(k)(1-\omega _l(k))}\,\left| 
\frac{d(1+k-\omega _l(k))}{dk}\right| ^{-1}\Big|_{k\rightarrow 0}=\frac 5{%
\eta ^2}+\frac 3\eta -\frac{353}{140}+O(\eta )\,,  \label{example3}
\end{eqnarray}
a divergent result too, but different from that of eq (\ref{example2}).
Which result should then one consider? The ambiguity comes from the fact
that the condition $k+1-\omega _l(k)=0$ is satisfied {\it only at }$k=0$; it
cannot be `approached from above', so to speak. Thus if one chooses to
regularize $I$ by defining: 
\begin{equation}
I_\eta =12\int_{\eta >0}^{+\infty }dk\int_{-\infty }^{+\infty }{\frac{%
d\omega _1}{\omega _1}}\int_{-\infty }^{+\infty }{\frac{d\omega _2}{\omega _2%
}}\,\delta (1-\omega _1-\omega _2)\,k^2|\omega _1|\,{\frak z}_l(k)\,\delta
(k^2-\omega _1^2)\,\delta (\omega _2-\omega _l(k))\,,  \label{example4}
\end{equation}
with $\eta \ll 1$ but kept different from zero until all intermediary steps
are performed, for both choices of eqs (\ref{example2}) and (\ref{example3})
one obtains: 
\begin{equation}
I_\eta =0\,,  \label{example5}
\end{equation}
and the ambiguity is lifted. This is the regularization procedure we adopt
in this work. The calculation becomes then mainly a matter of
straightforwardly disentangling the infrared-divergent pieces from the
finite ones. It very often necessitates an expansion for small $k$ of the
residue and cut functions using their definitions given in (\ref{ztl}) and (%
\ref{betatl}) respectively and some of their derivatives. Also, care must be
taken when handling the delta functions that occur, especially with their
first and second derivatives. Finally, extra care must be given to the order
in which the integrals over $k$ and $\omega $ are performed; it can be quite
subtle in some cases.

We now give the results for the different contributions to $a_{l0}^{(2)}$.
To ease the notation, we denote in a compact way: 
\begin{equation}
\int {\cal D}\equiv 9\int_\eta ^{+\infty }dk\,\int_{-\infty }^{+\infty }{%
\frac{d\omega _1}{\omega _1}}\,\int_{-\infty }^{+\infty }{\frac{d\omega _2}{%
\omega _2}}\,\delta (1-\omega _1-\omega _2)\,.  \label{integsign}
\end{equation}
We first have: 
\begin{equation}
\int {\cal D}\,{\frac{\omega _1}3}\left( 1-2\omega _1\right) \Theta
_1\,\partial _k\rho _{l2}=\frac 3{4\eta ^2}-\frac 9{40}-3\int_{1/2}^{+\infty
}dk\frac{1-2k}{1-k}\,\beta _l(1-k,k)\,.  \label{term1}
\end{equation}
Next we have: 
\begin{eqnarray}
\int {\cal D}\,\partial _k &&\Big[k^3\left( 1+\frac{k^2}3\right) \rho
_{l1}\rho _{l2}+\frac{2k}9\left( k^2(1-3k^2)+(1-4k^2)\omega _1\omega
_2-\omega _1^2\omega _2^2\right) \rho _{t1}\rho _{t2}  \nonumber \\
&&+{\frac{2\omega _1^2}3}\left( 1+{\frac{\omega _1}{k^2}}-{\frac{\omega _1^2%
}{k^2}}\right) \,\Theta _1\,\rho _{t2}\Big]=-\frac 34-\frac{40}{27\pi ^2\,}%
\,.  \label{term2}
\end{eqnarray}
It is worth mentioning at this stage that all contributions from $%
k\rightarrow +\infty $ vanish, indeed as it should be since only soft
contributions are to contribute. The next piece contains a term the residue
part of which we have discussed in the example above. It reads: 
\begin{eqnarray}
\int {\cal D}\,\frac{|\omega_1|}3\Big[4k^2\delta_1\rho_{l2}&&+2\omega_1
\delta_1\rho_{t2}+(1-2\omega_1)\big(1-{\frac{\omega _1^2}{k^2}}\big) %
\epsilon(\omega_1)\partial_k\Theta_1\rho_{t2}\Big]  \nonumber \\
&&=3\int_{1/2}^{+\infty }\frac{dk}{1-k}\left( 2k\,\beta _l(1-k,k)+\beta
_t(1-k,k)\right) \,.  \label{term3}
\end{eqnarray}

The next pieces are more delicate to handle: they need to work out
derivatives of delta functions in the presence of other delta functions with
different arguments. Nevertheless that can be done and we get: 
\begin{eqnarray}
\int {\cal D}\,\frac 16 &&\Big[2\omega _1k\Big( 1-\frac{\omega _1^2}{k^2}%
\Big) \,\partial _k\Theta _1\,\partial _k\rho _{t2}+{\frac{\omega _1}k}\Big( %
k^2-\omega _1^2\Big) \,\partial _k^2\Theta _1\,\rho _{t2}  \nonumber \\
&&-{\omega }_1\,\partial _k^2\,\Big( \Big( k-{\frac{\omega _1^2}k}\Big) %
\,\Theta _1\,\rho _{t2}\Big) \Big] =\frac 38-3\int_{1/2}^{+\infty }\frac{dk}{%
1-k}\beta _t(1-k,k)\,.  \label{term4}
\end{eqnarray}
The next term is worked out along the same lines as the previous one. We
have: 
\begin{equation}
\int {\cal D}\,\frac{\omega _1k}6\,\Theta _1\,\partial _k^2\rho _{l2}=-\frac 
9{8\eta ^2}+\frac{27}{80}+\frac 32\int_{1/2}^{+\infty }\frac{dk}{1-k}\left[
\beta _l(1-k,k)-k\,\partial _k\beta _l(\omega ,k)|_{\omega =1-k}\right] \,.
\label{term5}
\end{equation}

The next term is the most tedious: it necessitates the additional expansion
of $\partial _k\beta _t(\omega ,k)$ with $\omega =1-\omega _t(k)$ to order $%
k^2$. When this is done and the contributions put together, we get: 
\begin{equation}
\int {\cal D}\,\partial _k\left[ {\frac{k^2}9}\left( 1-2k^2-4(1-k^2)\omega
_1+6\omega _1^2-4\omega _1^3\right) \rho _{t1}\,\partial _k\rho _{t2}\right]
=\frac{125}{27\pi ^2\eta ^2}-\frac{1153}{189\pi ^2}-\frac{289536}{2187\pi ^4}%
\,.  \label{term6}
\end{equation}
The last term isn't more difficult. It reads: 
\begin{equation}
\int {\cal D}\,\left[ -{\frac 23}k^2\omega _1^2\,\epsilon (\omega
_1)\,\partial _{\omega _1^2}\delta _1\,\rho _{l2}\right] =-\frac 32%
\int_{1/2}^{+\infty }\frac{dk}{1-k}\,\left( \beta _l(1-k,k)+k\partial
_k\beta _l(\omega ,k)|_{\omega =1-k}\right) \,.  \label{term7}
\end{equation}

Putting all these contributions together we obtain: 
\begin{eqnarray}
a_{l0}^{(2)} &=&\left( -\frac 38+\frac{125}{27\pi ^2}\right) \frac 1{\eta ^2}%
-\frac{21}{80}-\frac{1433}{189\pi ^2}-\frac{289536}{2187\pi ^4}  \nonumber \\
&&\ \ -3\int_{1/2}^{+\infty }\frac{dk}{1-k}\left[ (1-4k)\beta
_l(1-k,k)+k\partial _k\beta _l(\omega ,k)|_{\omega =1-k}\right] \,.
\label{ffinal}
\end{eqnarray}
Note that the integral in the above equation (and all the other similar ones
actually) is finite. We can evaluate it numerically and we get: 
\begin{equation}
a_{l0}^{(2)}=\frac{0.09408}{\eta ^2}-4.45366\,.  \label{final}
\end{equation}
Hence, as announced, $a_{l0}^{(2)}$ is nonzero and quite divergent. Via eq (%
\ref{gamml}), this means $\gamma _l\left( 0\right) $ is itself
infrared-divergent. Indeed, using the value $a_{l0}^{(1)}=6.63538$, we get: 
\begin{equation}
\gamma _l\left( 0\right) =\frac{g^2N_cT}{24\pi }\left( \frac{0.09408}{\eta ^2%
}+2.18172\right) .  \label{gaml-final}
\end{equation}

We resume our initial discussion. Remember that the calculation of both $%
\gamma _t\left( 0\right) $ and $\gamma _l\left( 0\right) $ is performed
solely within the framework of the htl-resummation scheme. This latter
treats only the two scales $T$ and $gT$ and we have already remarked that
the scale $g^2T$ may play an important role in QCD\ at high $T$.\ This
suggests that the scheme as it is may not be robust enough at scales lower
than $gT$. One natural way this lack of robustness may manifest itself is
via an infrared sensitivity of the scheme when it comes to deal with
quantities that to lowest order are already of magnitude $g^nT$, $n\geq 2$.
We think the divergence in $\gamma _l\left( 0\right) $ we report in this
letter is a direct manifestation of this infrared sensitivity.

One may then ask why the result is finite for $\gamma _t\left( 0\right) $
whereas we find an infrared divergence for $\gamma _l\left( 0\right) $. It
is pertinent to note in this respect that there is a main difference between
the calculation in the htl scheme of $\gamma _l\left( 0\right) $ and that of 
$\gamma _t\left( 0\right) $: in order to get $\gamma _l\left( 0\right) $, we
already need an unavoidable expansion to order $p^2$ of the imaginary part
of the effective self-energy whereas for $\gamma _t(0)$ we do not \cite
{BPgamt,AAB}. As a matter of fact, we have determined the analytic
expression of the transverse-gluon damping rate $\gamma _t(p)$ to order $%
p^2. $ The coefficient $a_{t0}$, that of the zeroth order in $p^2,$ is
finite and as we said, equal to $a_{l0}^{(1)}$ of eq (\ref{al01}) \cite
{BPgamt}. But a preliminary investigation shows that the coefficient $a_{t1}$%
, that of order $p^2$, has contributing terms which are divergent in the
infrared \cite{AA}. It seems then that the expansion of the gluon effective
self-energies $^{*}\Pi $ of the htl scheme in powers of the soft momentum $p$
is infrared sensitive beyond zeroth order. $\gamma _l(0)$ necessitating such
an expansion to order $p^2$, hence an infrared-divergent result.

One may jump to conclude that the htl scheme is robust in the infrared, and
that all this simply means that the expansion in powers of the soft momentum
of the gluon effective self-energies is not valid in the first place. But
remember that this expansion {\it is necessary} for $\gamma _l(0)$; it is
`part of' its definition, so to speak. Since on physical grounds the damping
rates have to be finite for soft momenta, any consistent calculation scheme
must allow for a small-$p$ expansion with finite coefficients. If the
expansion yields infinite coefficients, it only means that the scheme in
which the calculation is performed needs to be cured and/or improved upon,
not that the principle of the expansion itself is invalid. It is our opinion
that this is the case for the htl scheme in the infrared sector. In other
words, an infrared-improved resummation scheme would still allow for the
same expansion but would remove the infrared sensitivity of the actual htl
scheme.

In fact, the infrared sector is not the first instance in which the htl
scheme yields infinities. Indeed, colinear singularities do appear when
external light-like momenta are involved and in response to this, the scheme
has been subjected to improvement in \cite{Fle-Rebh} where an improved
action that removes these singularities is proposed. We think the same
treatment is necessary in the infrared sector but the situation is more
problematic here. Indeed, such an improvement would very likely necessitate
the determination of the magnetic mass which is mostly believed to intervene
in a nonperturbative way. It may also be that the story does not end at the
scale $g^2T$ and new physics may manifest itself at lower scales $g^nT$ with 
$n>2$, though nothing is suggesting this for the moment. In any case, it is
pertinent to note that besides the result we obtain for $\gamma _l(0)$, what
we think is also interesting in this work is that it sets a calculable
framework in which the infrared divergences appear explicitly and by the
same token where to test an improved htl scheme when found. In the meantime,
we can have a very rough estimate of the scale at which the htl scheme
starts to be sensitive. If we demand that $\gamma _l(0)$ be equal to $\gamma
_t(0)$, we may set $a_{l0}^{(2)}$ equal to zero, which yields $\eta =0.145$
(in units of $m_g$). However, one should be prudent in considering this
value as an estimation, even rough, of the magnetic mass.

Another interesting point to discuss is the logarithmic behavior in $\gamma
(p)$ with $p$ soft but nonzero we mentioned in our introductory remarks \cite
{logdiv}. More precisely, one argues that $\gamma (p)\sim f(p)\ln (m_{{\rm %
mag}}/m_{{\rm elec}})$, where $f(p)$ is a well behaved function and the
ratio of the magnetic mass to the electric one is (at least) of order $g$.
Now as we suggested, our cut-off $\eta $ can be thought of representing the
magnetic scale, and thus in units of $m_{{\rm elec}}$ (denoted in our work
by $m_g$), it would also be of order $g$ at least. We notice then the
absence of $\ln \eta $ in the divergent behavior we obtain for $\gamma _l(0)$%
, see (\ref{gaml-final})$.$ This fact can be used to put back into question
the validity of the expansion we make. In particular, one may argue that a
term of the form $f(p)\ln p$ should be present in the expansion of the
effective self-energies $^{*}\Pi $, a term more in line with the logarithmic
behavior. First of all, such a sort of terms is ruled out by the nature of
the small$-p$ calculation itself and there is really no subtlety in this as
one can straightforwardly get convinced by following the steps reported in 
\cite{AAB}. Also, we stress that the htl scheme {\it must} allow for the
expansion in powers of $p$ and it actually {\it does}. What simply happens
is that the coefficients of the expansion it yields are infrared sensitive,
which means that the scheme is still yet not `complete' in this infrared
sector; in other words, it does not resum diagrams of order $g^2T$ and
smaller whereas our result suggests there is a need for such a further
resummation in one way or another. Once the scheme is properly improved
upon, then the new coefficients will cease to diverge.

But the point about the absence of $\ln \eta $ in the expression of $\gamma
_l(0)$ still remains. Indeed, if the $\ln \eta $ is present in the
expression of $\gamma (p)$, then one should expect it to remain present in
that of $\gamma (0)$. It is not difficult to understand its absence in the
transverse case. Indeed, when we expand $\gamma _t(p)$ to order $p^2$, the
coefficient $a_{t0}$ is finite but the coefficient $a_{t1}$ is
infrared-divergent, in particular the presence of $\ln \eta $ in it is not
ruled out from the outset. Therefore, when $p$ is small but different from
zero, the coefficients containing the infrared pieces dominate over the
finite ones like $a_{t0}$. But precisely at $p=0$, only the coefficient $%
a_{t0}$ survives and it is finite, hence the absence of $\ln \eta $ or any
other infrared-sensitive piece. However, the situation is more delicate in
the longitudinal case: this argument cannot be carried forward there since
already $a_{l0}$ is divergent and does not contain a $\ln \eta $ term.
Actually the longitudinal situation puts into light a problem we have not
raised thus far. Indeed, Besides $\ln \eta $, we expect to find power-like
divergences (some higher than $1/\eta ^2)$ in $a_{t1}$. These divergences
are more severe than $\ln \eta $ and yet have not been reported in the
literature. This issue is well beyond the scope of the present work and we
defer its discussion until we finish the disentanglement of all the infrared
pieces in $\gamma _t(p)$ to order $p^2$ \cite{AA}.

Finally, we have limited ourselves thus far to discussing the gluonic sector
only but certainly quarks play an important role too in the structure of hot
QCD. They themselves acquire a thermal mass $m_f$ to order $gT$ and their
damping rates start also at the scale $g^2T$, which means they are not a
priori shunned from infrared sensitivity.

\begin{acknowledgments}
We deeply thank Asmaa Abada for all her help and encouragements. O.A.
would also like to thank ICTP at Trieste for their kind hospitality where
some of the later part of this  work was carried. She had the opportunity
there to carry constructive discussions with Samina S. Masood.
\end{acknowledgments}

\end{document}